\documentclass[11pt,a4paper]{article}
\usepackage[utf8]{inputenc}
\usepackage[english]{babel}
\usepackage{amsmath}
\usepackage{amssymb}
\usepackage{amsfonts}
\usepackage{graphicx}
\usepackage[numbers]{natbib}

\DeclareMathOperator{\bin}{Bin}
\DeclareMathOperator{\mixbin}{MixBin}

\date{}
\title{Respondent-driven sampling and\\an unusual epidemic}
\author{Jens Malmros\thanks{Department of Mathematics, Stockholm university, SE-106 91 Stockholm, Sweden} \thanks{Corresponding author: jensm@math.su.se} \and Fredrik Liljeros\thanks{Department of Sociology, Stockholm university, SE-106 91 Stockholm, Sweden} \and Tom Britton\thanks{Department of Mathematics, Stockholm university, SE-106 91 Stockholm, Sweden}}

\begin{document}

\maketitle

\begin{abstract}

Respondent-driven sampling (RDS) is frequently used when sampling hard-to-reach and/or stigmatized communities. RDS utilizes a peer-driven recruitment mechanism where sampled individuals pass on participation coupons to at most $c$ of their acquaintances in the community ($c=3$ being a common choice), who then in turn pass on to their acquaintances if they choose to participate, and so on. This process of distributing coupons is shown to behave like a new Reed-Frost type network epidemic model, in which becoming infected corresponds to receiving a coupon. The difference from existing network epidemic models is that an infected individual can not infect (i.e.\ sample) all of its contacts, but only at most $c$ of them. We calculate $R_0$, the probability of a major ``outbreak'', and the relative size of a major outbreak in the limit of infinite population size and evaluate their adequacy in finite populations. We study the effect of varying $c$ and compare RDS to the corresponding usual epidemic models, i.e.\ the case of $c=\infty$. Our results suggest that the number of coupons has a large effect on RDS recruitment. Additionally, we use our findings to explain previous empirical observations.

\end{abstract}

\section{Introduction}

Hidden populations are groups of individuals which i) have strong privacy concerns due to illicit or stigmatized behaviour, and ii) lack a sampling frame, i.e., their size and composition are unknown. Examples of hidden populations include several groups that are at high risk for contracting and spreading HIV, e.g., men who have sex with men, sex workers, and injecting drug users~\cite{Beyrer2012,Kerrigan2012,Aceijas2004}; it is therefore of great importance to obtain reliable sampling methods for hidden populations in order to plan and evaluate interventions in the global HIV epidemic~\cite{Magnani2005,Lamptey2008}.

Respondent-driven sampling (RDS)~\cite{Heckathorn1997,Heckathorn2002} is a sampling methodology that utilizes the relationships between individuals in order to sample from the population. By combining an effective sampling scheme and the ability to produce unbiased population estimates, RDS has become the perhaps most preferred method when sampling from hidden populations. A typical RDS study starts with the selection of a group of seed individuals. Each seed is provided with a number of coupons, typically between three to five, to distribute to his or her peers in the population. An individual is eligible for participation upon presenting a coupon at the study site. Because recruitment takes place by coupons, participants remain anonymous throughout the study, but each coupon is numbered with a unique ID to keep track of who recruited whom. Incentives are given both for the participation of an individual as well as for the participation of those to whom he or she passed coupons. After participation, which commonly includes survey questions and possibly being tested for diseases, newly recruited individuals (i.e., respondents) are also given coupons to disperse among their contacts in the population. This procedure is then repeated until the desired sample size has been reached. The sampled individuals form a tree-like structure which is obtained from tracing the coupons. Recently, online based RDS methods (webRDS), where recruitment takes place via email and a survey is filled out at a designated web site, have also been put into use~\cite{Wejnert2008,Wejnert2009,Bengtsson2012}. There are several procedures available for estimating population characteristics from RDS data, most of which use a Markov model in order to approximate the actual recruitment process~\cite{Salganik2004,Volz2008,Gile2011JASA,Gile2011arXiv,LuEtal,malmros2013}; this is not the focus of the present paper.

A frequent problem in RDS studies is the inability of the recruitment process to reach the desired sample size due to premature failure of the recruitment chains started by the seeds~\cite{malekinejad2008}. This is often mitigated by additional seeds that enter the study as the rate of recruitment declines; e.g., in~\cite{malekinejad2008}, 43\% of reviewed RDS studies with available data reported that additional seeds were used. Relatedly, it has been observed in webRDS studies, where recruitment is allowed to go on until it stops by itself, that the recruitment process fails to reach a large proportion of the population despite additional seeds joining in at a later time~\cite{Bengtsson2012,Stein2014}. While there are most likely several reasons behind recruitment chain failure, such as community structure in the population causing chains to become stuck in a sub-network and/or clustering that has a similar effect, but more locally, an important reason is the limited number of coupons in the RDS recruitment process. This is the main focus of this paper. Furthermore, recruitment chain failure is highly associated with the ability of the recruitment process to start successful recruitment chains, the probability of such chains occurring, and the relative size of the population that is reached by an RDS study, all of which are related to quantities typically studied in epidemic modelling. As it turns out, it is possible to use models of infectious disease spread on social networks to describe coupon distribution in RDS, where the disease is defined as ``participation in the study'' and spreads by the RDS coupon distribution mechanism.

The simplest model of infectious disease spread is the Reed-Frost model, see e.g.~\cite[][p. 11-18]{andersson2000}, where in each generation $i$, each infectious individual independently infects each susceptible individual with the same probability. The individuals that were infected by the individuals in generation $i$ make up generation $i+1$ of infectious individuals in the epidemic. After spreading the disease, the individuals in generation $i$ are considered recovered (or dead) from the disease and are removed from the process. In the original version of the model, an infectious individual attempts to infect all susceptible individuals in the population. The model is however easily modified to the more realistic case when the structure of the population is described by a social network, hence imposing the restriction that an infectious individual only may spread the disease to his or her contacts in the social network independently of each other with the same probability. Infectious diseases are usually able to spread to all contacts of an individual, and consequently, the Reed-Frost model and other epidemic models defined on social networks do not impose any restrictions on the number of individuals that an infectious individual can infect other than those given by population structure. The RDS recruitment process differs from infectious diseases in that its spread is restricted by the limited number of coupons. Consequently, individuals with more population contacts than the number of coupons distributed to them have less capability of recruiting than if RDS recruitment were to spread in the usual manner of an epidemic, i.e.\ without any limitations. Depending on how the number of contacts (i.e, degrees) of population members are distributed, this may have a large effect on the capability of the RDS recruitment process to sustain and initiate recruitment. Furthermore, it may affect the ability of the recruitment process to reach a substantial proportion of the population, as the sampling procedure can limit recruitment to parts of the population. 

In this paper, we model RDS as an epidemic taking place on a social network by defining a Reed-Frost type model which has an upper limit on the number of individuals that an infectious individual could infect. We will use both infectious disease terminology and RDS terminology when referring to this model. In order to be able to specify the degree distribution of the social network, we use the configuration model~\cite{Molloy1995,Molloy1998} to describe the structure of the population. We calculate the \emph{basic reproduction number}, i.e., the number of individuals that are infected by a typical infectious individual during the early stages of the epidemic. This is often denoted by $R_0$. We say that there is a \emph{major outbreak} if a non-negligible proportion of the population is infected and calculate the probability $\tau$ of such outbreaks occurring. If $R_0\le1$, it is not possible for a major outbreak to occur, while if $R_0>1$, a major outbreak may occur. The critical value of $R_0=1$ is often referred to as the \emph{epidemic threshold}. We also calculate the relative size of an outbreak in case of a major outbreak $z$ using so-called \emph{susceptibility sets}~\citep{ball2001,ball2002}. Note that $\tau$ and $z$ are positive only if $R_0$ is larger than the epidemic threshold. We compare the RDS recruitment process to corresponding epidemics with unrestricted spread and investigate the effect of varying the number of coupons and the coupon transfer probability. To our knowledge, there are no previous studies of epidemics on networks that describes behaviour similar to the present one, although the model in \cite{martin1986} allows for a restriction on the number of individuals that an infectious individual can infect in a homogeneously mixing population (i.e.\ a population without network structure). 

\section{Models}

\subsection{Network model}\label{Subsec:NetworkModel}

We consider a configuration model network consisting of $n$ vertices. In later calculations, we will assume that $n\to\infty$. Each
individual $i, i=1,\ldots,n,$ is assigned an i.i.d.\ number of stubs (half-edges) $d_i$ from a prescribed distribution
$D$ having support on the non-negative integers. The network is then formed by pairing stubs together uniformly at
random. If $\sum_{i=1}^nd_i$ is odd, an edge is added to the $n$:th vertex (this does not influence our results in the
limit of infinite population size). This construction  allows the formation of multiple edges and self-loops; it is
however well known that the fraction of these is small if $D$ has finite second moment. Specifically, the probability of
the resulting graph being simple is bounded away from 0 as $n\to\infty$; see \citep[Theorem 7.8]{hofstad2009} and
\citep[Lemma 5.3]{britton2007}. Hence we can condition on the graph being simple given that $E(D^2)<\infty$.
Alternatively, we may proceed by removing multiple edges and self-loops from the generated graph since asymptotically
this does not change the degree distribution if $D$ has finite second moment; see \citep[Theorem 7.9]{hofstad2009}.
Hence, we will from now on assume that the resulting graph is simple. Moreover, the graph is locally tree-like when
$E(D^2)<\infty$, meaning that it with high probability does not contain short cycles \citep{britton2007}. Hence, we can
take advantage of the branching process \citep[e.g.,][]{athreya2011} approximations that are often used for epidemics, see e.g.~\citep[][ch. 3]{andersson2000}. In what follows, we will assume that the degree distributions considered have finite second moment.

\subsection{Epidemic model}\label{Subsec:EpidemicModel}

On this graph, describing the social structure in a community, we define an epidemic model mimicking the RDS recruitment process. In this model, becoming infected corresponds to participating in the RDS study. Initially, all members of the population (vertices) are susceptible. The epidemic starts with one randomly selected individual (vertex), the index case, being infected from the outside. The infected individual uniformly selects $c$ of his or her neighbours in the population and infects them
independently of each other with the same probability $p$. The parameter $c$ corresponds to the number of coupons in RDS and the parameter $p$ to the probability of being successfully recruited to the RDS study. If the infected individual has less than $c$ contacts, he or she infects all his or her contacts independently of each other with probability $p$. The newly infected individuals
make up the first generation of the epidemic. After spreading the disease, the initially infected individual recovers
and becomes immune (or dies) and has no further role in the epidemic. The individuals in the first generation each in
turn select $c$ of their neighbours excluding the one who infected them (which for the first generation is the index
case), regardless of whether they are susceptible or not. If an individual has less than $c$ neighbours excluding the
one who infected him or her, he or she selects all of his or her neighbours. Then, they infect the selected contacts
that are susceptible, independently of each other with probability $p$, and then recover; contacts with already infected individuals have no effect. The now infected individuals
form the second generation of the epidemic. The disease continues to spread in the same fashion from the
second generation and onward until there are no newly infected individuals in a generation. The individuals that were
infected during the course of the epidemic make up the outbreak, and the number of ultimately infected individuals is
the final size of the outbreak. Note that if we let $c=\infty$, we get the standard Reed-Frost epidemic taking place on the configuration model network~\cite{britton2007}. 

Because an individual only tries to infect those he or she selected, the spread of the disease, or coupon distribution mechanism, in our model is more similar to that of webRDS than physical RDS. We discuss this further and present other possible coupon distribution mechanisms in Section~\ref{Sec:Discussion}.

\section{Calculations}\label{Sec:Calculations}

\subsection{The basic reproduction number $R_0$}\label{Subsec:R0}

Assume that we have a configuration model graph $G$ of size $n$, where $n$ is large, and let the degree distribution of $G$ be
$D$, where $P(D=k)=p_k$. The degree of a given neighbour of an individual follow the \emph{size-biased} degree distribution $\tilde D$, where $P(\tilde
D=k)=\tilde p_k=kp_k/E(D)$. Assume that we have an epidemic spreading on this graph according to the description in
Subsection \ref{Subsec:EpidemicModel}. The degree of the index case is then distributed as $D$, and the degree of
infected individuals in later generations during the early stages of an outbreak is distributed as $\tilde D$. As previously mentioned in Subsection \ref{Subsec:NetworkModel}, the graphs generated by the configuration model will with high probability not contain short cycles, meaning that we can approximate the spread of the epidemic with a
(forward) branching process. Let $X$ and $\tilde X$ be the offspring of the ancestor (i.e., the index case) and of the
later generations in this branching process, respectively. Given that the index case has degree $k\le
c$, he or she can at most infect $k$ neighbours. If the index case has degree larger than or equal to $c+1$, he
or she infects at most $c$ neighbours. Because infections happens independently with the same probability $p$, we
have that, conditionally on the degree, the probability that the index case infects $j$ neighbours is
\begin{equation}\label{Eq:CondNumberInfectedIndex}
P(X=j|D=k)=\binom{c\wedge k}{j}p^j(1-p)^{(c\wedge k)-j},
\end{equation}
where $j=0,\ldots,c\wedge k$. Infectious individuals in later generations have one less contact available for
infection (the one that infected them). Hence, we get that, conditionally on the degree, the probability that an
infectious individual in later generations infects $j$ neighbours is
\begin{equation}\label{Eq:CondNumberInfectedLaterGen}
P(\tilde X=j|\tilde D=k)=\binom{c\wedge (k-1)}{j}p^j(1-p)^{(c\wedge (k-1))-j},
\end{equation}
where $j=0,\ldots,c\wedge (k-1)$.

Because the ability of an individual to spread the disease will depend on its degree, the offspring distributions are obtained by conditioning on
the degree: 
\begin{align}
P(X=j) &=\sum_{k=j}^\infty P(X=j|D=k)p_k;\\
P(\tilde X=j) &=\sum_{k=j+1}^\infty P(\tilde X=j|\tilde D=k)\tilde p_k,
\end{align}
where $j=0,\ldots,c$, and the probabilities $P(X=j|D=K)$ and $P(\tilde X=j|\tilde D=k)$ come
from Eqs.~\eqref{Eq:CondNumberInfectedIndex} and \eqref{Eq:CondNumberInfectedLaterGen}, respectively. From standard
branching process theory~\cite{athreya2011} we have that $R_0$ is the expected number of individuals
that get infected by an infectious individual in the second and later generations; hence
\begin{align}\label{Eq:R0}
R_0 &=E(\tilde X) = \sum_{j=0}^c j\sum_{k=1}^\infty P(\tilde X=j|\tilde D=k)\tilde p_k\\
&=\sum_{j=0}^c j\left(\sum_{k=j}^{c-1} \binom{k}{j}p^j(1-p)^{k-j}\tilde p_k +
\binom{c}{j}p^j(1-p)^{c-j}\left(1-\sum_{k=1}^c\tilde p_k\right)\right).\notag
\end{align}
The obtained $R_0$ is increasing in $p$ and $c$, and for a fixed $p$, $R_0\to R_0^{\rm (unrestricted)}$ as $c\to\infty$, where $R_0^{\rm (unrestricted)}$ is the $R_0$ value for the standard Reed-Frost epidemic on a configuration model network, given by \citep{britton2007}
\[R_0^{\rm (unrestricted)}=\left(E(D)+\frac{\text{Var}(D)-E(D)}{E(D)}\right).\]

\subsection{Probability of major outbreak}\label{Subsec:ProbMaj}

When $R_0>1$, it is possible for a major outbreak to occur. The probability $\tau$ of such an outbreak occurring is given
by the survival probability of the approximating branching process, which we get by standard techniques. We first
consider a branching process with offspring distribution $\tilde X$ for all individuals, i.e.\ also for the index case.
Let the extinction probability of this process be $\tilde\pi$. For the process to die out, all the branching processes
initiated by the offspring of the ancestor must die out; hence by conditioning on the number of offspring in the first
generation of the process, we get
\begin{equation}\label{Eq:pitilde}
\tilde\pi =\sum_{j=0}^c\tilde\pi^j P(\tilde X=j)=\tilde\rho(\tilde\pi),
\end{equation}
where $\tilde\rho$ is the probability generating function of $\tilde X$. The solution to Equation~\eqref{Eq:pitilde} is obtained numerically. In our original branching process the ancestor has
offspring distribution $X$ and later generations have offspring distribution $\tilde X$. Again by conditioning on the
number of individuals in the first generation, we get that the extinction probability $\pi$ of the original branching
process is
\begin{align}\label{Eq:pi}
\pi &=\rho(\tilde\pi),
\end{align}
where $\tilde \pi$ is the solution to Equation~\eqref{Eq:pitilde} and $\rho$ is the probability generating function of $X$. The solution to Equation~\eqref{Eq:pi} is given
by numerical calculations, and we obtain the probability of a major outbreak $\tau=1-\pi$.

Note that if we have $1<s<\infty$ initially infected individuals in the epidemic, the probability of a major outbreak is $1-\pi^s$, which approaches 1 as $s$ becomes large. The number of initially infected individuals does not affect $R_0$ or the relative size of a major outbreak calculated in Subsection~\ref{Subsec:RelativeSize}.

\subsection{Relative size of a major outbreak}\label{Subsec:RelativeSize}

The relative size of a major outbreak in case of a major outbreak $z$ can be obtained using susceptibility sets, constructed as follows. For each individual $i$, we can obtain a random list of which
neighbours that $i$ would infect given that it were to be infected. By combining the lists from all individuals in the
population, it is possible to construct a directed graph with all vertices (individuals) in which there is an arc from
vertex $i$ to vertex $j$ if $j$ is in $i$:s list. The susceptibility set of an individual $j$ consists of all
individuals in this directed graph, including $j$ itself, from which there is a directed path to $j$. Hence, $j$:s
susceptibility set is such that the infection of any individual in the set would result in the ultimate infection of
$j$. Note that $j$ will be infected in the epidemic if and only if the initially infected individual is in $j$:s susceptibility set. 

The susceptibility set of a randomly chosen individual, $i_0$ say, can be approximated with a (backward) branching process in which $i_0$ is the only member of the zeroth generation. We consider the number of neighbours that, if they were to be infected, would infect $i_0$ (as opposed to previously when we considered the number of neighbours that an individual would infect were it to be infected). Suppose that $i_0$ have degree $d$. Because all neighbours of $i_0$ contact him or her with the same probability $\theta$ independently of each other, the number of neighbours that contact him or her is $\bin(d,\theta)$-distributed; hence, the unconditional distribution of the number of neighbours that contact him or her is a mixed binomial distribution with parameters $D$ and $\theta$. We now derive an equation for the contact probability $\theta$. The degree distribution of the neighbouring individuals is $\tilde D$, so we obtain
\begin{equation}
\theta =\sum_{k=0}^\infty\theta_k\tilde p_k,
\end{equation}
where $\theta_k$ is the probability that a neighbour with degree $k$ contacts $i_0$. Because a neighbour of $i_0$ with degree
$k$ has to be contacted first in order to become infected, only $k-1$ edges are available for him or her to spread the
disease. Therefore, a neighbour must have at least degree two in order to first become infected and then contact
$i_0$. If a neighbour has degree $k\ge c+2$, he or she first selects $c$ of the available $k-1$ contacts and then attempts to spread the disease to them. Hence, the contact probabilities are 
\begin{equation}
\theta_k=\left\{
\begin{array}{ll}
0, &k=0,1;\\
p, &k=2,\ldots,c+1;\\
\frac{c}{k-1}p, &k=c+2,c+3,\ldots.
\end{array}\right.
\end{equation}
The probability that a neighbour makes contact with $i_0$ depends on his or her degree. Hence, the degree distribution
of individuals in the first generation, i.e.\ those neighbours of $i_0$ that makes contact with $i_0$, and of individuals in later generations in the backward branching process is altered by the fact that they have contacted another individual. Conditionally on the event that a contact has been made, call it $C$, the distribution of the degree $D^*$ of an
individual in the first and later generations of the susceptibility set process is given by 
\begin{align}
P(D^*=k) &= P(\tilde D=k|C)\notag\\
&=\frac{P(C|\tilde D=k)P(\tilde D=k)}{\sum_{k=0}^\infty P(C|\tilde D=k)P(\tilde D=k)}\notag\\
&= \frac{\theta_k\tilde p_k}{\theta},
\end{align}
so
\begin{equation}
P(D^*=k)=\left\{
\begin{array}{ll}
0,&k=0,1;\\
\frac{p\tilde p_k}{\theta}, &k=2,\ldots,c+1;\\
\frac{cp\tilde p_k}{(k-1)\theta}, &k=c+2,c+3,\ldots.
\end{array}\right.
\end{equation}
An individual in later generations of the process will be contacted by any of his or her neighbours independently of other neighbours with
the same probability $\theta$. Given that this individual has degree $k$, the number of neighbours that contact him or her is binomially distributed with parameters $k-1$ and $\theta$. Hence, the unconditional distribution of the number of neighbours that contact an individual in later generations is mixed binomial with parameters $D^*-1$ and $\theta$. 

If the approximating backward branching process contains few individuals, it is unlikely that $i_0$ will be infected, whereas if the process reaches a large number of individuals (i.e.\ grows infinitely large), there is a positive probability that $i_0$ will not escape infection. More specifically, the probability that $i_0$ will be infected during a major outbreak is given by the survival probability of the backward branching process. Because $i_0$ is chosen randomly, we also have that the relative size of an outbreak in case of a major outbreak is given by the survival probability of the backward branching process. Let $Y$ be the number of offspring of the ancestor and $Y^*$ the number of offspring of individuals in later generations in the approximating branching process, respectively. Hence, $Y\sim\mixbin(D,\theta)$ and $Y^*\sim\mixbin(D^*-1,\theta)$. We obtain the survival probability of the process similarly as in Subsection~\ref{Subsec:ProbMaj}. Let the extinction probability of a branching process with offspring distribution $Y^*$ be $\pi^*$. We have
\begin{align}\label{Eq:pistar}
\pi^* &= \sum_{j=0}^\infty(\pi^*)^jP(Y^*=j) = E((\pi^*)^{Y^*})\notag\\
&= E(E((\pi^*)^{Y^*}|D^*)) = E(1-\theta+\theta\pi^*)^{D^*-1}\notag\\
&=\sum_{k=2}^\infty(1-\theta+\theta\pi^*)^{k-1}P(D^*=k)\notag\\
&=\frac{p}{\theta}\sum_{k=2}^{c+1}(1-\theta+\theta\pi^*)^{k-1}\tilde p_k+\frac{cp}{\theta}\sum_{k=c+2}
^\infty(1-\theta+\theta\pi^*)^{k-1}\frac{\tilde p_k}{k-1}.
\end{align}
The solution to Equation~\eqref{Eq:pistar} for $\pi^*$ is obtained numerically. Let the extinction probability of the approximating branching process be $\pi'$. Then,
\begin{align}\label{Eq:piprime}
\pi' &= \sum_{j=0}^\infty(\pi^*)^jP(Y=j) = E((\pi^*)^Y)\notag\\
&= E(E((\pi^*)^Y|D)) = E(1-\theta+\theta\pi^*)^D\notag\\
&= f_D(1-\theta+\theta\pi^*),
\end{align}
where $f_D(\cdot)$ is the probability generating function of $D$ and $\pi^*$ is the solution to Equation~\eqref{Eq:pistar}. The solutions to Equation \eqref{Eq:piprime} is obtained numerically, and the relative final size of the epidemic in case of a major outbreak is $z=1-\pi'$.

A rigorous proof of that $z=1-\pi'$ is beyond the scope of this paper. It has been proved that for Reed-Frost epidemics on random intersection graphs~\citep{ball2014} and Reed-Frost epidemics on configuration model graphs~\citep{ball2013} that the proportion of infected during the epidemic converges in probability to the survival probability of the backward branching process. Similar arguments could also be used for our process to provide a formal proof. Additionally, we believe that the techniques described in~\cite{barbour2013} could be used to obtain stronger results for the whole epidemic process. 

\section{Numerical results and simulations}\label{Sec:Results}

We now numerically examine the analytical results obtained in Section~\ref{Sec:Calculations}. In particular, we examine the relation between $R_0$, $\tau$, and $z$ and the parameters $c$ and $p$, and compare the RDS recruitment process with unrestricted epidemics. We use two different degree distributions in our calculations, the Poisson degree distribution and a variant of the power-law degree distribution with exponential cut-off given by $p_k\propto k^{-\alpha}\exp(-k/\kappa)$, $k=1,2,\ldots$, where $\alpha$ is the power-law exponent and $\kappa$ refers to the exponential cut-off~\cite[e.g.][]{newman2002epidemics}.

\begin{figure}
\centering
\includegraphics[width=6cm]{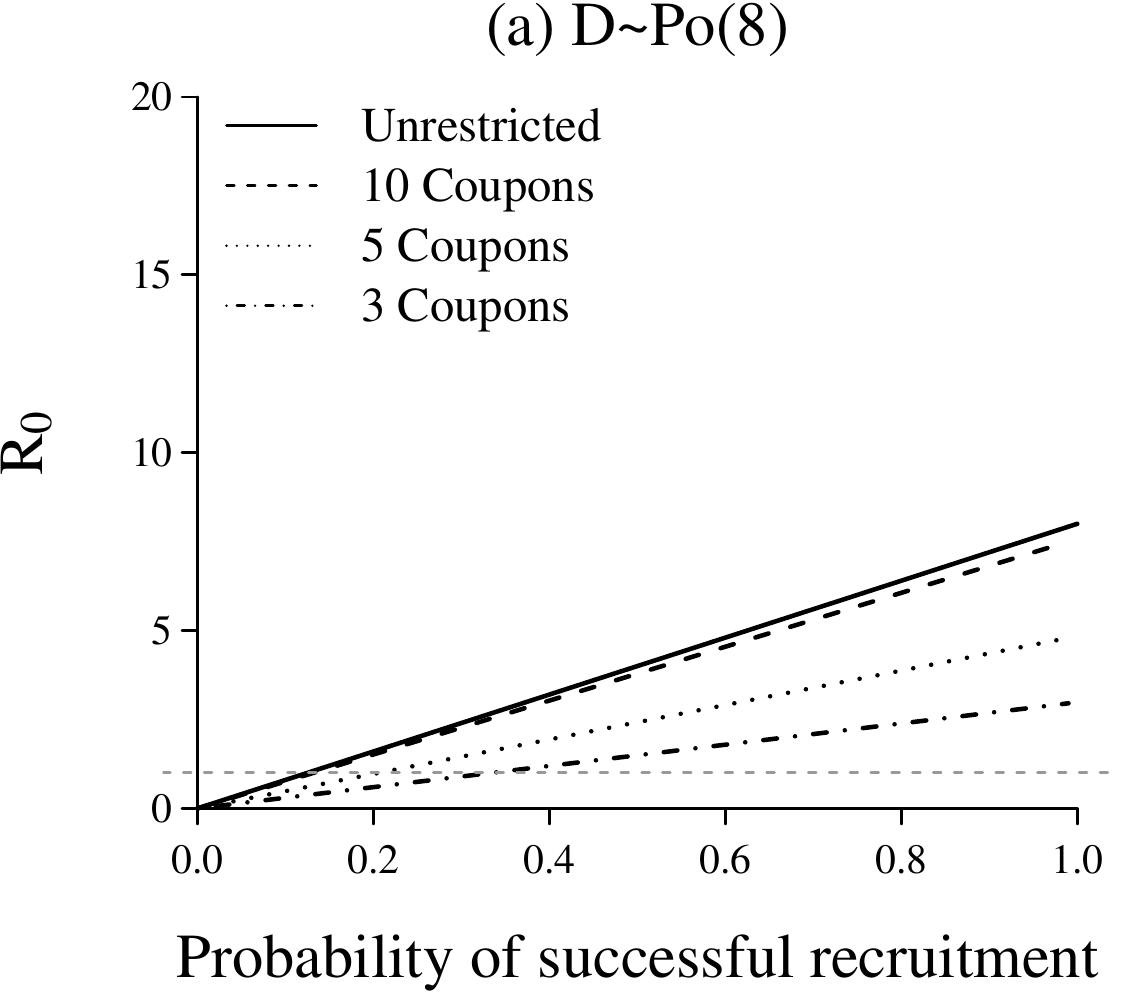}
\includegraphics[width=6cm]{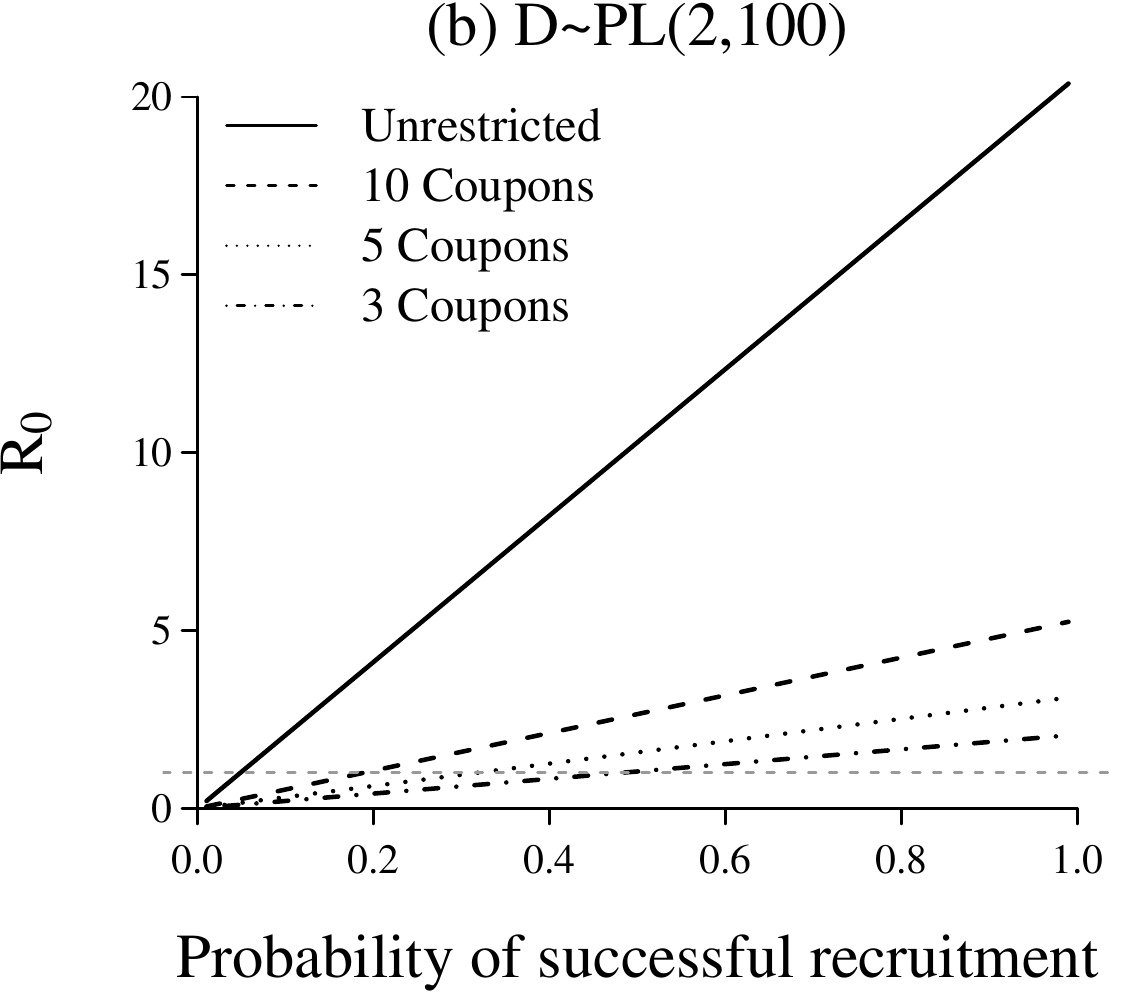}
\caption{Comparison of $R_0$ for unrestricted epidemics and RDS recruitment processes with 10, 5, and 3 coupons and $p\in[0,1]$. Plot (a) show the results for the Poisson degree distribution with parameter $\lambda=8$ and plots (b) show the results for the power-law degree distribution with parameters $\alpha=2$ and $\kappa=100$. The dashed horizontal lines shows the threshold value $R_0=1$.}\label{Fig:R0}
\end{figure}

In Figure~\ref{Fig:R0}, we show the $R_0$ values for the RDS recruitment process with $c=3,5,10$ and the unrestricted epidemic for $p\in[0,1]$. Figure~\ref{Fig:R0}~(a) shows the results for the Poisson degree distribution with parameter $\lambda=8$ and Figure~\ref{Fig:R0}~(b) shows the results for the power-law degree distribution with parameters $\alpha=2$ and $\kappa=100$. For both degree distributions and a fixed value of $p$, the limitation imposed by the number of coupons on disease spread yields smaller $R_0$ values for the RDS recruitment process when compared to the unrestricted epidemic for all values of $c$. Especially for the power-law degree distribution, all values of $c$ give much smaller $R_0$ values than those of the unrestricted epidemic, and the value of $p$ for which $R_0$ becomes larger than 1 (i.e., the epidemic threshold) is larger than that of the unrestricted epidemic for all values of $c$.

\begin{figure}
\centering
\includegraphics[width=6cm]{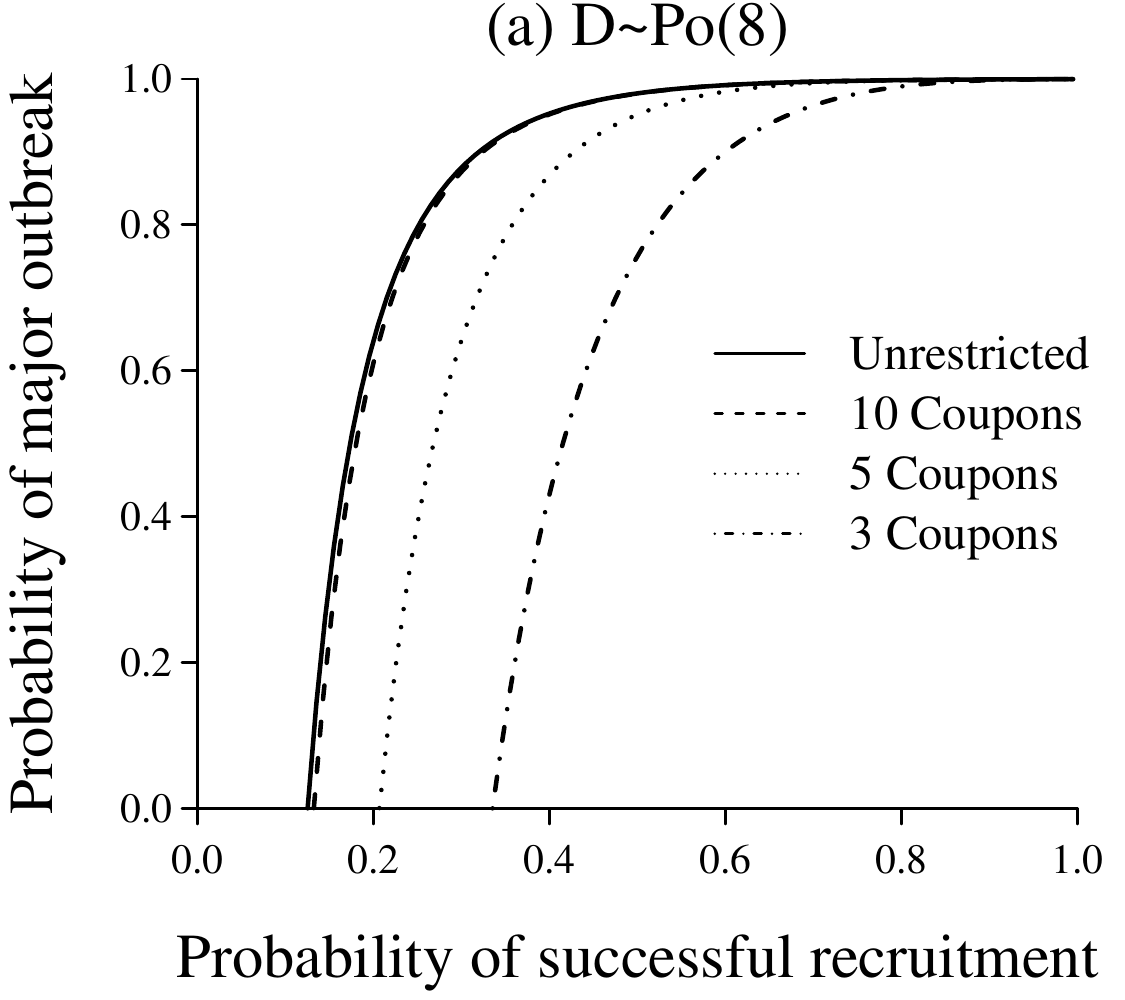}
\includegraphics[width=6cm]{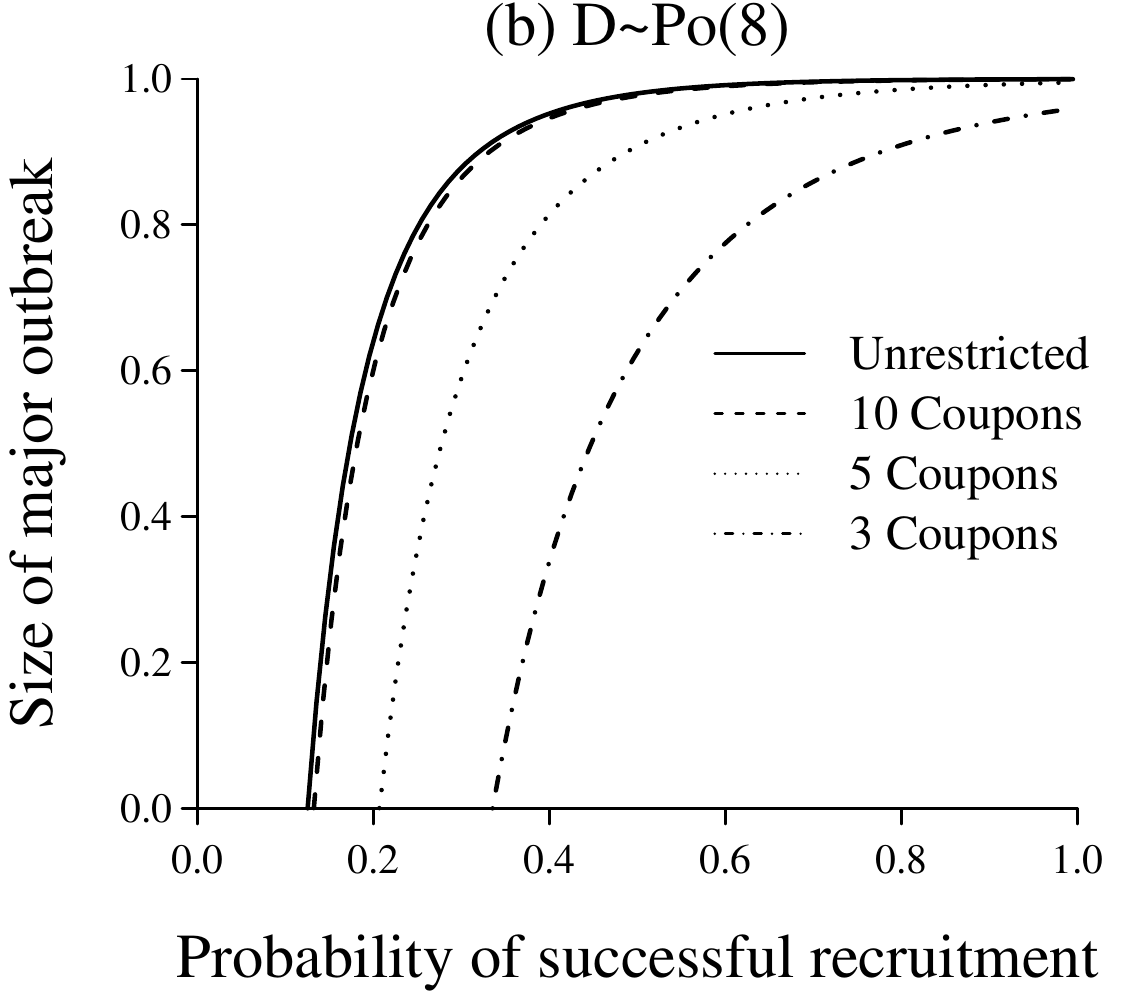}\vspace{10pt}
\includegraphics[width=6cm]{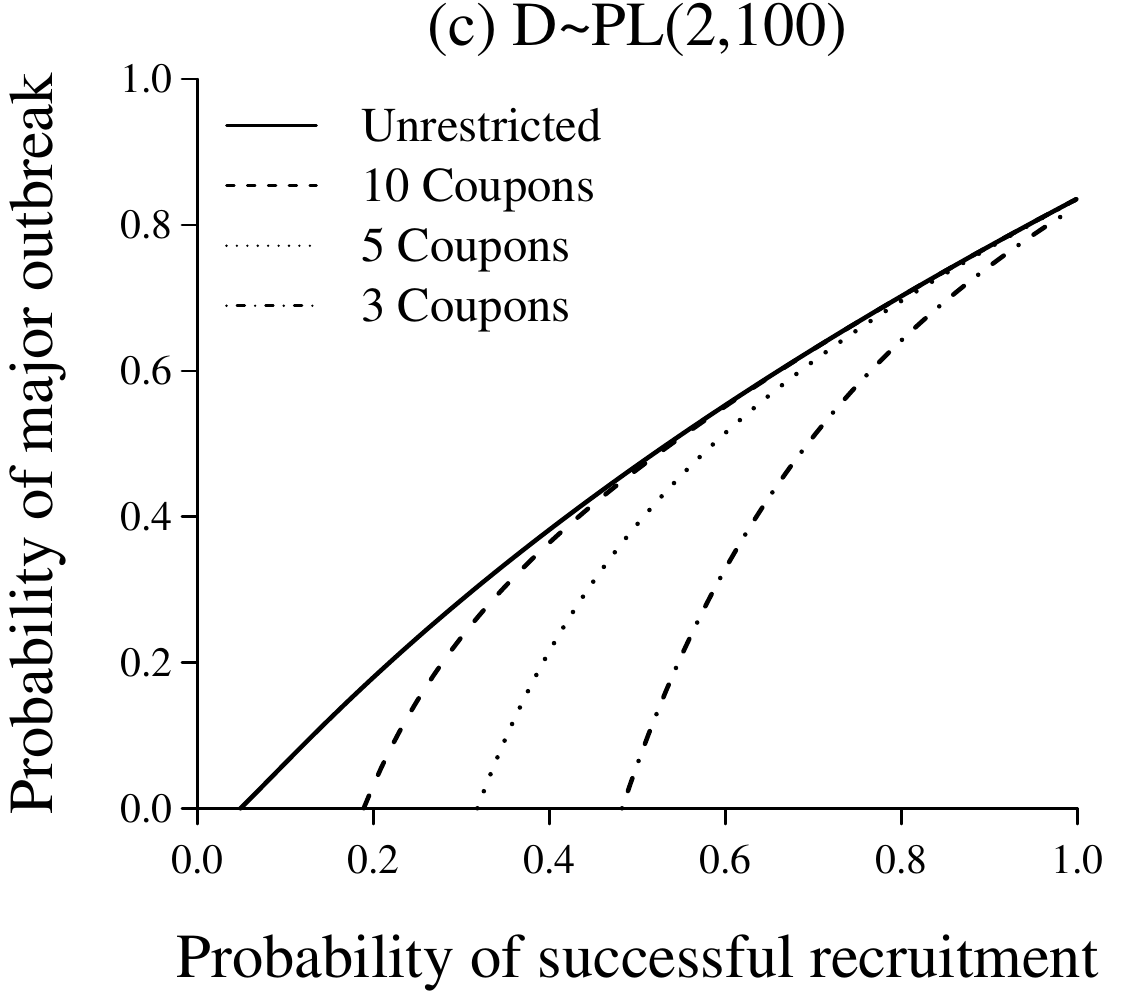}
\includegraphics[width=6cm]{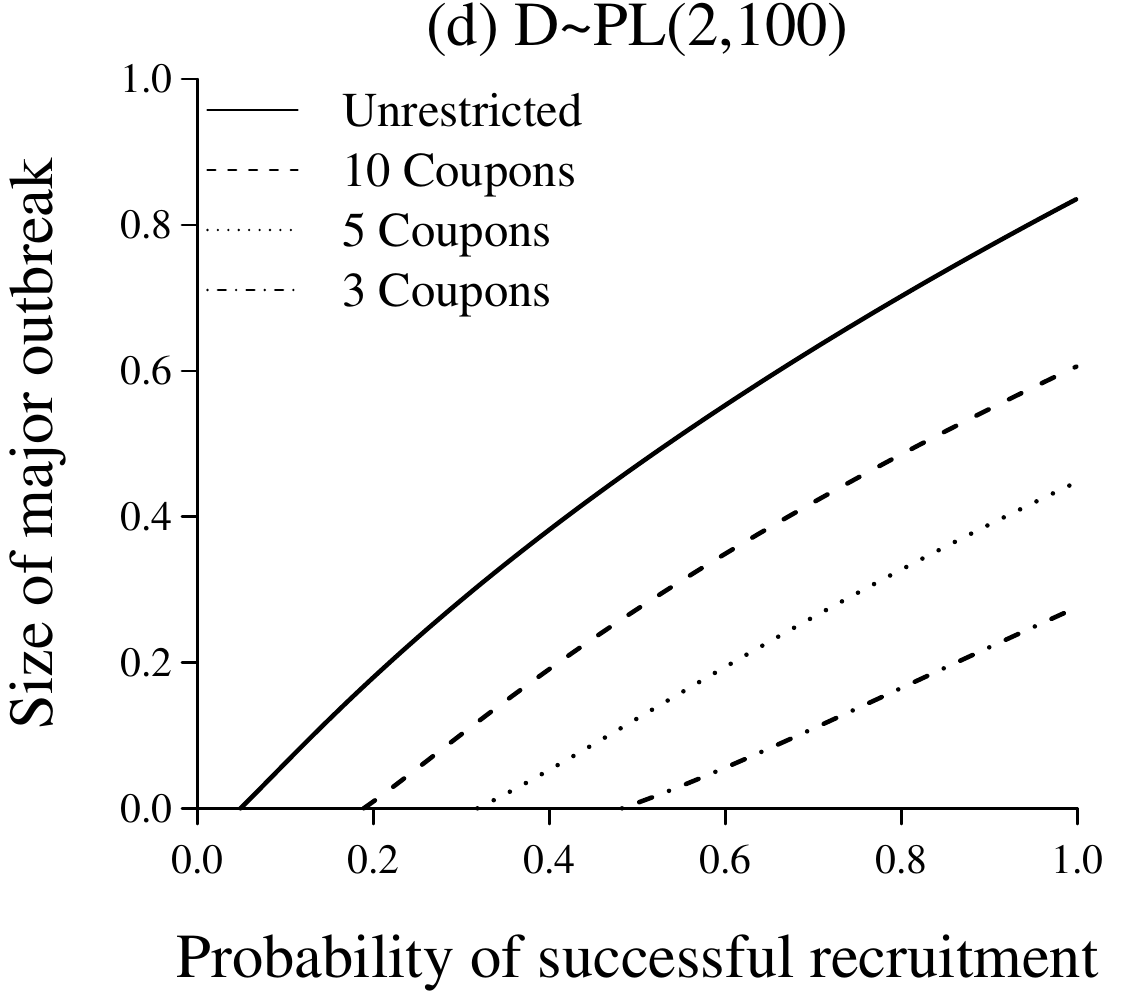}
\caption{Comparison of the asymptotic probability of a major outbreak and relative size of a major outbreak for
unrestricted epidemics and RDS recruitment processes with 10, 5, and 3 coupons and $p\in[0,1]$. Plots (a) and (b) show
the results for the Poisson degree distribution with parameter $\lambda=8$ and plots (c) and (d) show the results for
the power-law degree distribution with parameters $\alpha=2$ and $\kappa=100$.}\label{Fig:ProbMajSizeOutbreak}
\end{figure}

Figure~\ref{Fig:ProbMajSizeOutbreak} shows the values of $\tau$ and $z$ for the RDS recruitment process with $c=3,5,10$ and the unrestricted epidemic for $p\in[0,1]$. Figures~\ref{Fig:ProbMajSizeOutbreak}~(a) and \ref{Fig:ProbMajSizeOutbreak}~(b) show the results for $\tau$ and $z$, respectively, for the Poisson degree distribution with parameter $\lambda=8$ and Figures~\ref{Fig:ProbMajSizeOutbreak}~(c) and \ref{Fig:ProbMajSizeOutbreak}~(d) show the results for $\tau$ and $z$, respectively, for the power-law degree distribution with parameters $\alpha=2$ and $\kappa=100$. The relative size of a major outbreak is always smaller than the probability of a major outbreak for both degree distributions. For both degree distributions, the probability of a major outbreak for the RDS recruitment process is smaller than that of the unrestricted epidemic for small values of $p$ and approaches that of the unrestricted epidemic when $p\to1$. For the power-law degree distribution, the size of a major outbreak is much smaller than that of the unrestricted epidemic for all values of $c$ and $p$. 

\begin{figure}
\centering
\includegraphics[width=6cm]{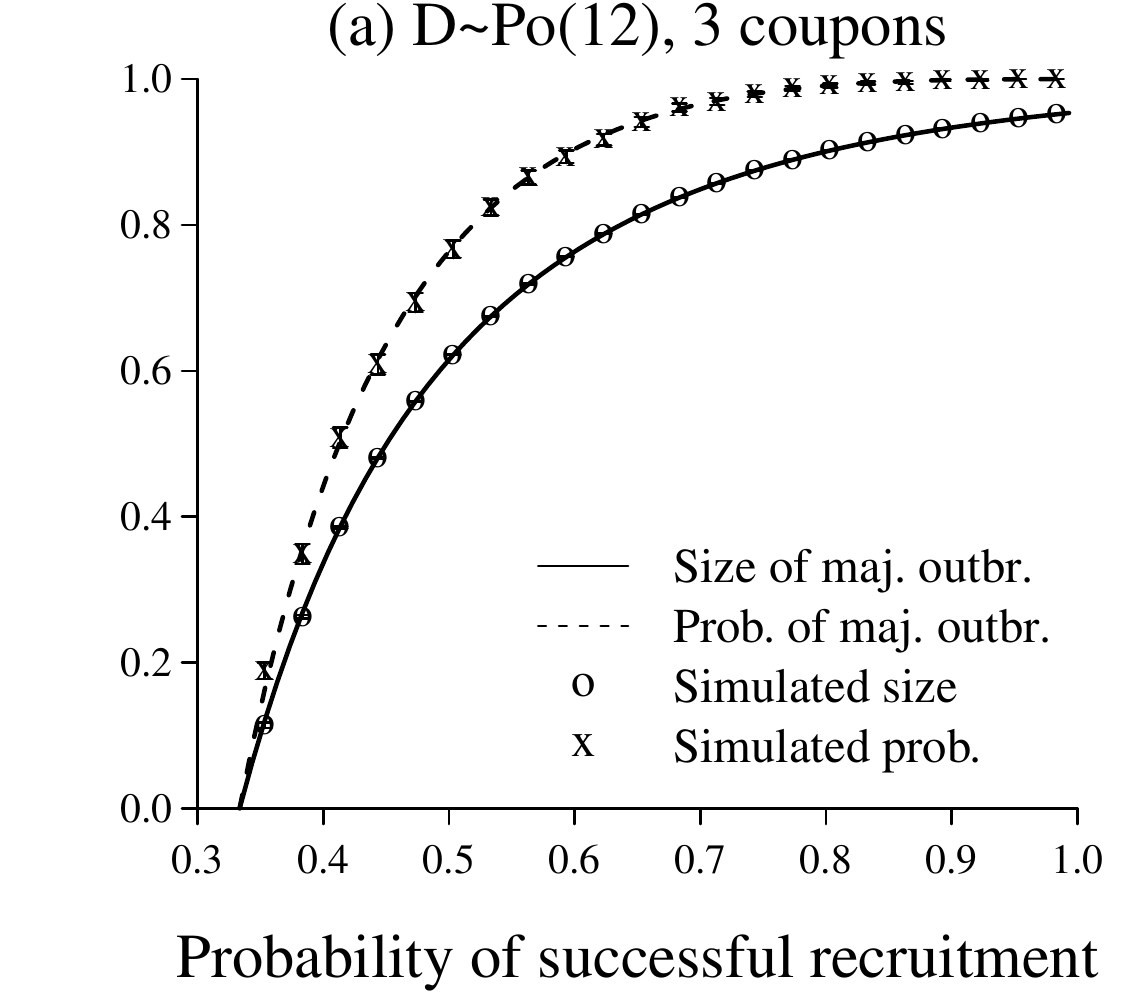}
\includegraphics[width=6cm]{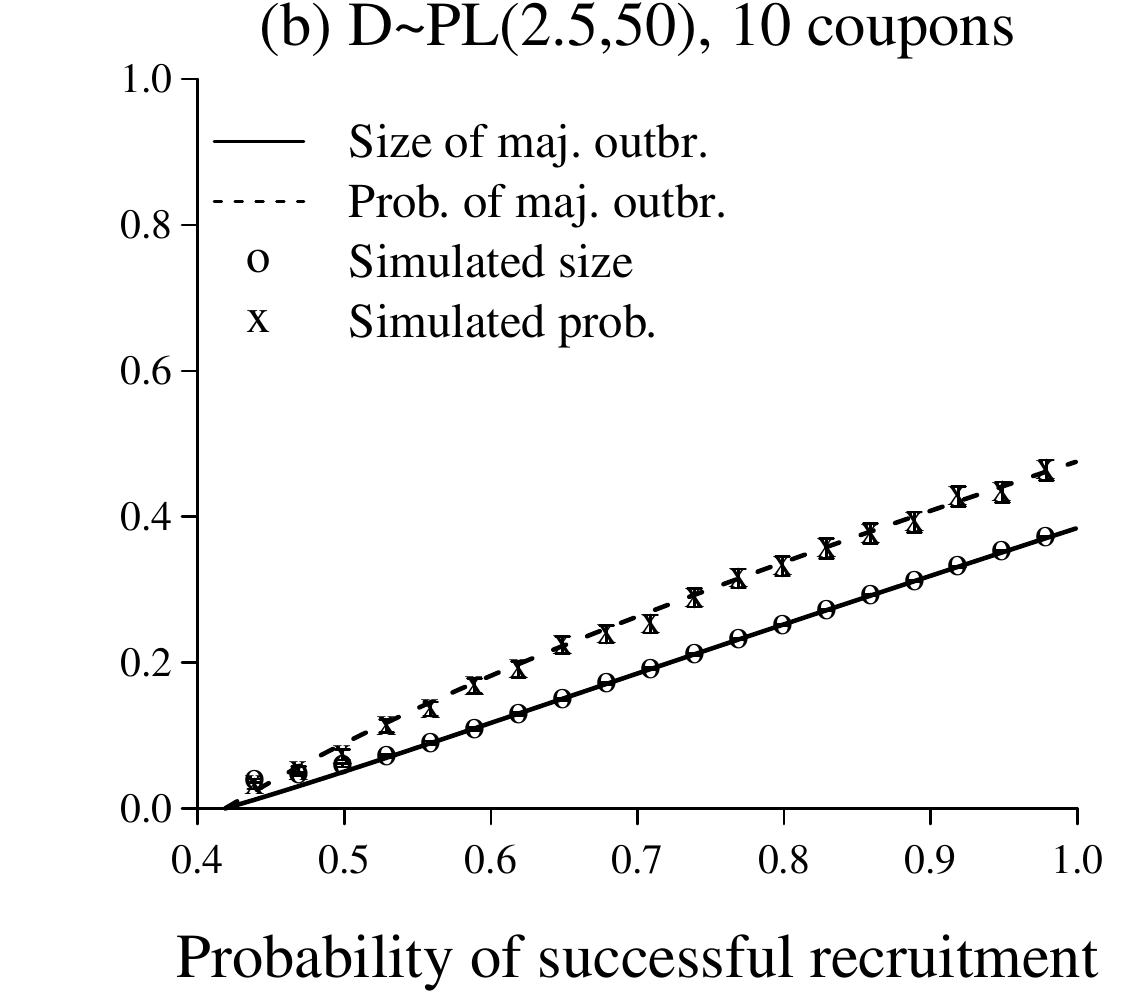}
\caption{Comparison of results from simulations of RDS recruitment processes and the asymptotic probability and relative size of a major outbreak.
Plot (a) shows the results for the Poisson degree distribution with parameter $\lambda=12$ for processes with $c=3$ and plot (b) shows the
results for the power-law degree distribution with parameters $\alpha=2.5$ and $\kappa=50$ for processes with $c=10$. Note that the error bars for the simulated relative size are very narrow and not visible for most simulated values. Also note that the horizontal scales are different.}\label{Fig:Simulations}
\end{figure}

We also make a brief evaluation of the adequacy of our asymptotic results in finite populations by means of
simulations. From simulated RDS recruitment processes (as described by the model), we estimate the probability of a major outbreak and the relative
size of a major outbreak in case of a major outbreak by the relative proportion of major outbreaks and the mean relative size of major outbreaks, respectively. Given a degree distribution and number of coupons $c$, let $p_c$ be the smallest value of $p$ for which the process is above the epidemic threshold. Each simulation run consists of generating a network of size 5000 by an erased configuration model approach \citep{britton2006}, for which we make use of the iGraph R package~\citep{igraph}. Then, RDS recruitment processes are run on the generated network for values of $p\in[p_c,1]$. In Figure~\ref{Fig:Simulations}, we show the estimated probability of a major outbreak $\hat\tau$ and estimated relative size of a major outbreak in case of a major outbreak $\hat z$ for varying $p$ and the corresponding asymptotic results. Figure~\ref{Fig:Simulations} (a) shows the results for the Poisson degree distribution with parameter $\lambda=12$ from 5000 simulations runs of RDS recruitment processes with 3 coupons. Figure~\ref{Fig:Simulations} (b) shows the results for the power-law degree distribution with parameters $\alpha=2.5$ and $\kappa=50$ from 5000 simulation runs of RDS recruitment processes with 10 coupons. In both Figures~\ref{Fig:Simulations} (a) and (b), we show error bars for the estimates based on $\pm2$ standard errors, where the standard error for $\hat\tau$ is estimated as $SE(\hat\tau)=(\hat\tau(1-\hat\tau)/m))^{1/2}$, where $m$ is the number of simulations, and the standard error for $\hat z$ is estimated as $SE(\hat z)=(\hat\sigma^2/m_{maj})^{1/2}$, where $\hat\sigma^2$ is the sample variance of the relative final sizes of major outbreaks and $m_{maj}$ is the number of simulations resulting in a major outbreak. 

Note that it is not well defined what constitutes a major outbreak in small, finite populations. Usually, the threshold for when an outbreak constitutes a major outbreak is determined by inspecting the distribution of outbreak sizes. Typically, this distribution is bimodal with modes at 0 and $z$, corresponding to small and major outbreaks. In our model, outbreak sizes will depend on $p$. For $p$ close to $p_c$, where ``close'' depends on the degree distribution, small and major outbreaks are indistinguishable. Consequently, it is difficult to estimate $\tau$ and $z$ for such values of $p$. In Figure~\ref{Fig:Simulations}, we have chosen to set the (relatively small) threshold for major outbreaks to 2\% of the population over the whole interval $[p_c,1]$. This yields fairly correct estimates for $p$ close to $p_c$ and does not affect estimates for $p$ further away from $p_c$.  

We see that both the estimated probability of a major outbreak and the estimated relative size of major outbreak in case
of a major outbreak are very well approximated by the asymptotic results for both the evaluated degree distributions. As
pointed out in \citep{ball2009}, the relative size of the epidemic is more efficiently estimated than the probability of
a major outbreak because each simulation yields many (correlated) observations of the backward process and only one
observation of the forward process.

\section{Discussion and conclusions}\label{Sec:Discussion}

When the RDS recruitment process is compared to the corresponding unrestricted epidemic, it is clear that the limited number of coupons has a large impact on $R_0$ and the value of $p_c$ corresponding to the epidemic threshold, the probability of a major outbreak, and the relative size of a major outbreak in case of a major outbreak. This is especially true for the power-law degree distribution, for which in particular $R_0$ and $z$ is much smaller than for the corresponding unrestricted epidemic. In social networks with power-law degree distribution, the vast majority of individuals will have small degrees. For these individuals, the probability of being infected in an epidemic will be small. Also, such an individual will, once infected, have few or no contacts to spread the disease to. Hence, the spread of an epidemic in such networks will be highly dependent on a few individuals with very large degrees that have the capacity to infect many of their (small degree) neighbours. Because of the relatively small value of $c$, the potential of large degree individuals to spread the disease is much impaired in RDS compared to an unrestricted epidemic with the same $p$, hence impairing the spread of the epidemic as a whole. 

The impact of the number of coupons on the RDS recruitment process may in part explain why some RDS studies experience difficulties in obtaining the desired sample size and/or recruiting a substantial proportion of the study population. Given $p$, the number of coupons will be crucial to whether $R_0$ is above or below the epidemic threshold for the recruitment process; in the latter case all recruitment chains will eventually fail. Moreover, the proportion of the population recruited by the RDS recruitment process may be small even given that $p$ is relatively large and a major outbreak occurs. For some parameter combinations, the proportion reached can be very small; this is especially important to consider in webRDS. We illustrate this by considering the webRDS studies in~\cite{Bengtsson2012} and~\cite{Stein2014}. In both studies, each respondent were allowed to make 4 recruitments. In the latter study, 66\% of started recruitment chains had a depth of one generation (i.e.\ index case and one generation of recruitments) and 11\% had a depth of three generations or more. This indicates that $R_0$ is below the epidemic threshold for this study and therefore, recruitment never takes off. In the former study, the majority of recruitments come from long recruitment chains, implying that $R_0$ is above the epidemic threshold. Still, recruitment eventually declined and stopped completely before reaching a large part of the population despite additional seeds joining the study. As we see in Section~\ref{Sec:Results} however, relatively many parameter combinations with $R_0>1$ yields small $z$ values, which could explain the observed behaviour. For both studies, heterogeneity in network structure, such that, locally $R_0<1$, may also be an explanation. It would be of interest to find proper inference procedures for our model to be used in further evaluation of actual RDS studies with respect to the quantities studied in this paper.

One might consider other ways to distribute coupons. The coupon distribution mechanism in our model, where a respondent selects some of his or her neighbours for attempted coupon transfer while ignoring those neighbours that were not selected, is most similar to a webRDS process. In a physical RDS study where coupons are handed over from person to person, a respondent may attempt to distribute a coupon to another neighbour if the originally intended recipient declines (here, distributing a coupon implies study participation). This modified mechanism is given as follows. A respondent first attempts to give a coupon to a randomly chosen neighbour. If the coupon is rejected, the respondent may try to distribute the same coupon to another neighbour, randomly chosen among those who previously have not been offered a coupon. When the coupon is accepted, the procedure is repeated starting by randomly selecting among those neighbours that have not been offered a coupon. When there are no more neighbours and/or coupons left, no further distribution attempts are made. The offspring probabilities in the branching process are the same as previously for individuals with degree less than the number of coupons, but the distribution of the number of coupons given out by an individual with degree larger than $c$ will be tilted towards larger values compared to the previous model. The probabilities in Eq.~\eqref{Eq:CondNumberInfectedIndex} now become
\begin{equation}
P(X=j|D=k)=\left\{
\begin{array}{ll}
\binom{k}{j}p^j(1-p)^{k-j},&j<c;\\
\sum_{i=c}^k\binom{k}{i}p^i(1-p)^{k-i},&j=c.
\end{array}\right.
\end{equation}

\begin{figure}
\centering
\includegraphics[width=6cm]{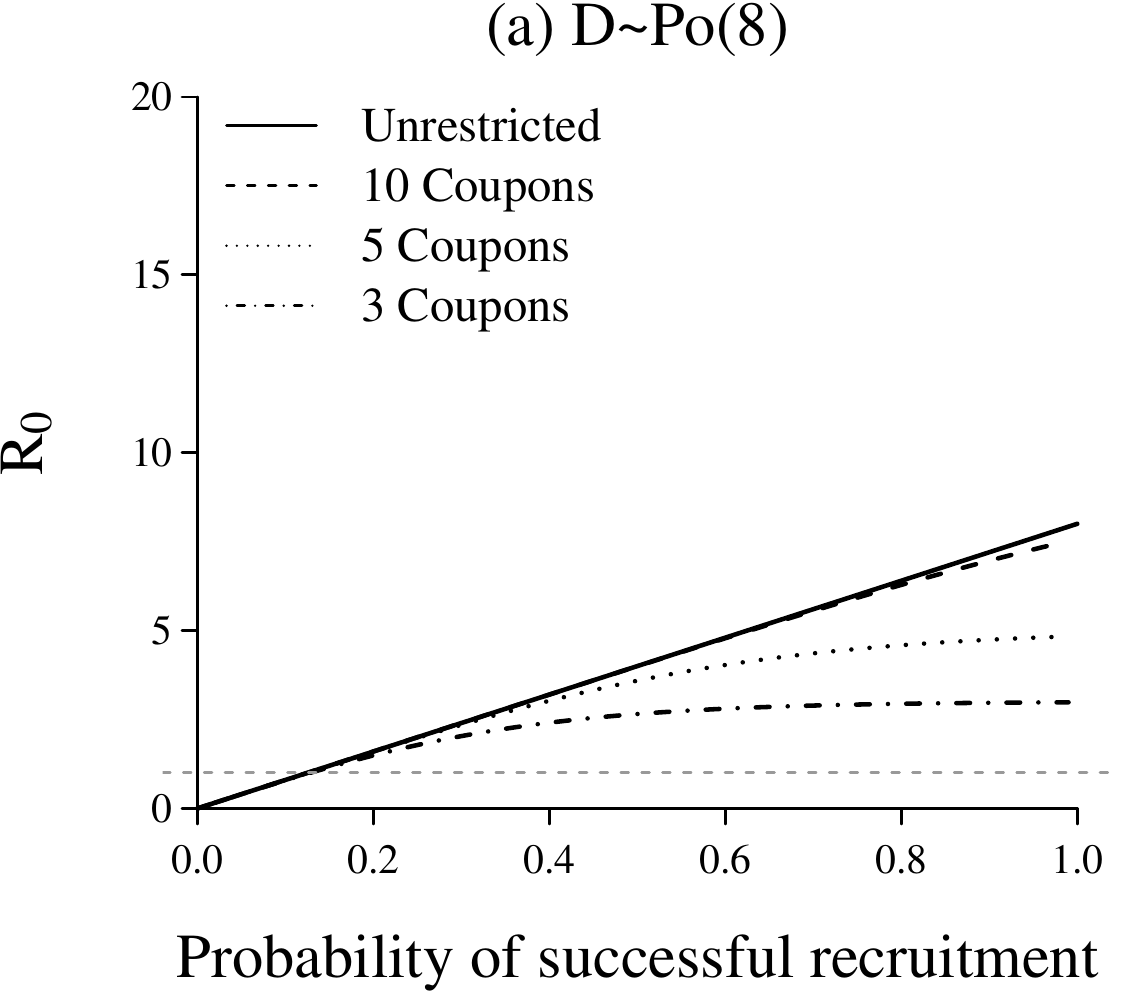}
\includegraphics[width=6cm]{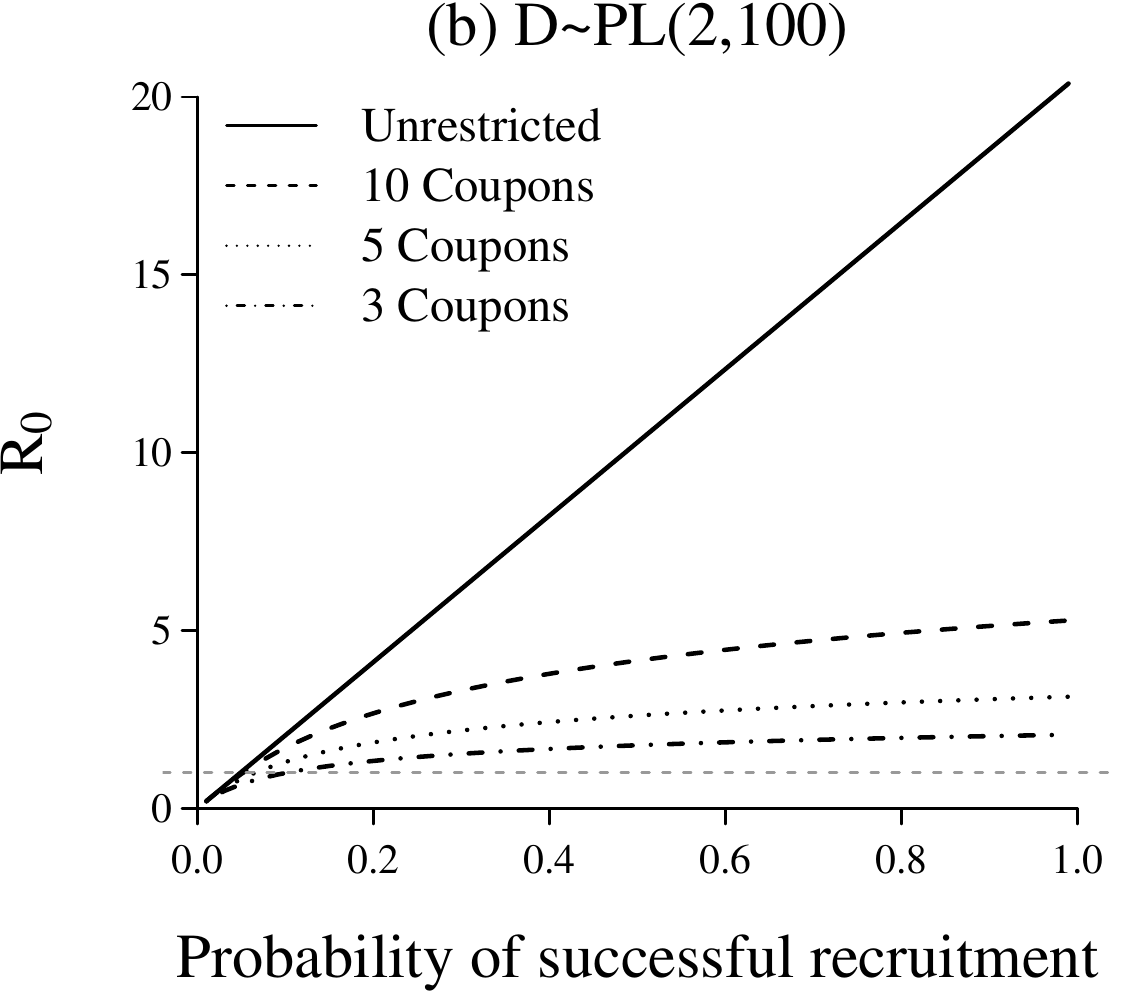}
\caption{Comparison of $R_0$ for unrestricted epidemics and RDS recruitment processes where a recruiter tries to distribute a coupon until success. Plot (a) shows the results for the Poisson degree distribution with parameter $\lambda=8$ and plot (b) shows the results for the power-law degree distribution with parameters $\alpha=2$ and $\kappa=100$.}\label{Fig:R0ModifiedRDS}
\end{figure}

It is straightforward to calculate $R_0$ and $\tau$ using the same techniques as in Sections~\ref{Subsec:R0} and~\ref{Subsec:ProbMaj}. Figure~\ref{Fig:R0ModifiedRDS} shows the $R_0$ values for the modified RDS recruitment process with $c=3,5,10$ and the unrestricted epidemic for $p\in[0,1]$. In Figure~\ref{Fig:R0ModifiedRDS}~(a), we show the results for the Poisson degree distribution with $\lambda=8$ and in Figure~\ref{Fig:R0ModifiedRDS}~(b) we show the results for the power-law degree distribution with parameters $\alpha=2$ and $\kappa=100$. It is clear that $R_0$ is larger for the modified recruitment process for all $p$ compared to the process described in Subsection~\ref{Subsec:EpidemicModel} and the $p$ value corresponding to the epidemic threshold is considerably smaller. When $p\to1$, the $R_0$ values converges to those seen in Figure~\ref{Fig:R0}. 
Because the modified process has similar epidemic threshold values in terms of $p$ for different $c$, the corresponding $\tau$ values (not shown) are close to those for the unrestricted epidemic when $R_0>1$. For the final size of the epidemic, the calculations are much harder to derive and is thus out of the scope of this paper. There are several other complications that could be considered in terms of coupon distribution. E.g., it is not likely that all coupon distribution attempts of a respondent will have the same success probability, both because the respondent may act differently depending on how many attempts he or she has previously made and because the relations to his or her neighbours may be different. Other complications include different respondent behaviour depending on (measurable) individual characteristics, geographical variations, and time dependence. 

Overall, our results indicate that RDS studies which experience difficulties with respect to recruitment chain failure could benefit from an increased number of coupons, which would reduce the number of additional seeds needed. Furthermore, the longer recruitment chains obtained as a result of an increased number of coupons are more likely to reach remote parts of the population and meet equilibrium criteria for inference. As the recruitment potential of RDS increases from an increased number of coupons, the time to reach the desired sample size is shortened. Additionally, the study time is not subject to unexpected prolongation due to the addition of seeds. Hence, an increased number of coupons may result in lower and more predictable study costs. For webRDS studies in particular, the increase in the proportion of the population reached due to increasing the number of coupons facilitates larger sample sizes. We therefore advise that the recruitment potential of a planned RDS study should be considered beforehand so that the number of coupons could be chosen large enough to facilitate sustained recruitment and an acceptable sample size. Other factors may also increase recruitment potential. The coupon transfer probability $p$ could be increased by e.g.\ larger incentives or improved information about the study; this has an immediate effect on $R_0$, $\tau$, and $z$. Additionally, the selection of seeds could also affect recruitment capability, see e.g.\ \cite{wylie2013} where different seed selection methods produce very different recruitment scenarios. In general, it is of interest to further study why certain RDS studies are more successful in reaching the desired sample size with a modest number of seeds.

The presented epidemic model is a novel contribution to the area of stochastic epidemic models and although many results from Reed-Frost epidemics on configuration model networks are expected to hold for this model, several properties of it remain to be studied. There are a number of extensions that can be considered, e.g.\ different recruitment probabilities through unequally weighted edges,  controlling for network structural properties, e.g.\ clustering, and modifying the coupon distribution mechanism as previously described. 

\section*{Acknowledgements}
J.M. was supported by the Swedish Research Council, grant no. 2009-5759. T.B. and F.L. are grateful to Riksbankens jubileumsfond (contract
P12-0705:1) for financial support. 

\bibliographystyle{JRSI}

\end{document}